\documentclass[fleqn,usenatbib]{mnras}
\usepackage{fix-cm}

\usepackage{newtxtext,newtxmath}
\usepackage{makecell}

\usepackage[T1]{fontenc}

\DeclareRobustCommand{\VAN}[3]{#2}
\let\VANthebibliography\thebibliography
\def\thebibliography{\DeclareRobustCommand{\VAN}[3]{##3}\VANthebibliography}

\usepackage{graphicx}	
\usepackage{amsmath}	
\usepackage{upquote}

\title[Polarization in GX 9+1]{First spectro-polarimetric study of the neutron star low-mass X-ray binary GX 9+1}

\author[V. Prakash et al.]{
V. P. Shyam Prakash,$^{1,2}$\thanks{E-mail: shyamvp151@gmail.com}
Vivek K. Agrawal,$^{1}$
and A. M. Vinodkumar$^{2}$
\\
$^{1}$Space Astronomy Group, ISITE Campus, U. R. Rao Satellite Center, ISRO, Bengaluru 560037, India.\\
$^{2}$Department of Physics, University of Calicut, Kerala, 673635, India.
}

\date{Accepted XXX. Received YYY; in original form ZZZ}

\pubyear{\the\year{}}

\begin{document}
\label{firstpage}
\pagerange{\pageref{firstpage}--\pageref{lastpage}}
\maketitle

\begin{abstract}
We present the first spectro-polarimetric study of the bright atoll source GX 9+1, using the simultaneous Imaging X-ray Polarimetry Explorer (\textit{IXPE}), and Neutron star Interior Composition Explorer (\textit{NICER}) observations. 
The source was observed to remain in the soft state, with no changes in state throughout the observation period.
The source does not show significant polarization in the 2-8 keV energy range. However, a significant polarization (3.3$\sigma$) was detected in the 2-3 keV range, with a polarization degree of 3.3 $\pm$ 0.8\% and a polarization angle of 11 $\pm$ 7\textdegree.
We used the simultaneous energy spectra from \textit{NICER} (0.6 - 11 keV) and \textit{IXPE} (2-8 keV) to study the spectral properties of the source during observations. 
The observed spectrum of the source can be well described by a combination of Comptonized blackbody emission from the neutron star surface (\textit{compbb} model in \texttt{XSPEC}) and thermal Comptonized component with seed photons from the accretion disc.
The spectral properties of GX 9+1 during the observation are consistent with those of other bright atoll-sources in the soft state.
However, the high polarization degree observed in the low-energy band does not align with previous \textit{IXPE} observations of other atoll-sources. 
This observed polarization in the source is attributed to the strong polarization of the Comptonized blackbody component. 
We discuss the results from the spectro-polarimetric studies in the context of various accretion disc and coronal geometries of the source.

\end{abstract}




\section{Introduction}\label{intro}
Weakly magnetized neutron stars (NS) in low-mass X-ray binaries (NS-LMXBs) are bright X-ray sources that accrete matter from the companion star via Roche lobe overflow (\citealt{2006csxs.book..623T}).
NS-LMXBs are classified as `Z' and `atoll' sources based on the pattern they trace in the hardness-intensity diagram (HID) and color-color diagram (CCD; \citealt{1989A&A...225...79H}). 
The spectral and timing behaviors are strongly correlated with the position in both Z and atoll tracks \citep{1989A&A...225...79H, annurev.aa.27.090189.002505}.
While a Z-source \big(with luminosity, $L \sim 10^{37} - 2\times 10^{38} \, \mathrm{erg\,s^{-1}}$ \big) traces three branches (horizontal branch (HB), normal branch (NB) and flaring branch (FB)) in the CCD \citep{2006csxs.book...39V, 1994A&A...289..795K},
atoll-sources \big($L \sim 10^{36} - 10^{37} \, \mathrm{erg\,s^{-1}}$\big) are observed in the high soft state (HSS) or the low hard state (LHS). They display island and banana states in the CCD. (\citealt{1995xrbi.nasa..252V}).
Some atoll-sources have also shown Z-shaped tracks in the CCD (\citealt{2002ApJ...576..391B}).
Sources such as IGR J17480–2446 (\citealt{2010ATel.2952....1A}), XTE J1701-462 (\citealt{Homan_2010}), XTE J1806-246, Cir X-1 (\citealt{Homan_2010}) have shown the behaviors of both the Z and atoll-sources.
The transient X-ray source, XTE J1701-462, has shown evolution from a Cyg-like Z-source to a Sco-like Z-source and eventually transforms into an atoll-source during the 2006-2007 outburst (\citealt{2009ApJ...696.1257L}).
Also, the transient source, XTE J1806–246 has exhibited the characteristics of Z-sources at higher luminosities and atoll-properties at lower luminosities (\citealt{Wijnands_1999}).

X-ray spectral studies provide insight in regrading the emission mechanisms of NS-LXMB sources.
The X-ray spectra of NS-LMXBs are modeled using a combination of soft thermal component, which dominate below 1 keV, and a hard component resulting from the Comptonization of seed photons by the hot plasma. 
Different approaches have been adopted for modeling the X-ray spectra of NS-LMXBs in the past.
In one approach, the soft thermal component is generated from a multi-temperature accretion disk, while the hard component arises due to the Comptonization of seed photons from the NS (\citealt{1984PASJ...36..741M}).
This could be from the boundary layer (BL; \citealt{2001ApJ...547..355P}) or spreading layer (SL) near the neutron star surface.
The presence of a spherical shell close the neutron star surface was proposed by \citealt{2001ApJ...555L..45T} to describe the normal branch oscillations (NBOs) in Z-sources.
This approach has been extensively used to describe the X-ray spectra of various NS-LMXBs in the past \citep{Barret_2000, 10.1111/j.1365-2966.2003.07147.x, Agrawal:2022jbm}. 
Another approach assumes the softer component as a single temperature blackbody from the NS surface, while the hard component is generated by the Comptonized disc photons (\citealt{1988ApJ...324..363W}). 
There were also attempts to explain the continuum emission in NS-LMXBs using two thermal components. One from the multi-color accretion disc, while the other from the spreading and/or boundary layer \citep{Lin_2010, 2012A&A...543A..50D, 10.1093/mnras/stad1606}. 
The presence of reflection features in the X-ray spectrum have been detected in several sources in addition to the primary continuum emission \citep{2008ApJ...674..415C,Ludlam_2017, 2009MNRAS.398.2022D, 2017MNRAS.466.4991M}. 

Modeling the X-ray spectra of NS-LMXB sources help to shed light on the nature of the emission from the accretion disc and corona. 
However, the spectroscopic models are degenerate with respect to the geometry and location of the corona. 
X-ray polarization measurements can help to constrain the geometry of corona in such systems. 
With the launch of the Imaging X-ray Polarimetry Explorer (IXPE; \citealt{2022JATIS...8b6002W}) in December 2021, it is now possible to obtain such measurements in addition to spectroscopic and timing information.
Since its launch in 2021, \textit{IXPE} has observed several LMXB sources which includes both `Z', `atolls' and peculiar-sources \citep{2023ApJ...943..129C, 2023A&A...676A..20U, 2023MNRAS.521L..74C, 2023ApJ...953L..22D, 2024A&A...692A.123G}.
\textit{IXPE} observations of most of the atoll-sources were conducted in the soft state and Z-sources have been observed in different branches (HB, NB and FB).
In Z-sources, a high value of Polarization degree (PD) (up to 4–5\%) is observed in the HB, while a low PD is observed in NB and FB \citep{2023A&A...674L..10C, 2024A&A...684A.137F, Jayasurya:2023udz}. 
For all the atoll-sources observed so far using \textit{IXPE}, (GS 1826-238 (\citealt{Capitanio_2023}), GX 9+9 \citep{Ursini2023, 2023MNRAS.521L..74C} and 4U 1820-303 (\citealt{DiMarco_2023})), a low value of PD is observed in 2–8 keV (see review by \citealt{2024Galax..12...43U}).
However, for the dipping source 4U 1624-49 (\citealt{Saade_2024}), a high degree of polarization (PD = 3.1 $\pm$ 0.7\%) is observed.
Energy-dependent polarization behavior has been observed in many of these sources. 
In atoll-sources, an increase in PD with respect to energy is generally observed in the soft state (\citealt{2024Galax..12...43U}).
The polarization signal is related to the hard component, characterized by a higher polarization compared to the soft thermal emission from the accretion disc \citep{2023MNRAS.521L..74C,2024A&A...690A.230G, 2023MNRAS.519.3681F, 2023A&A...674L..10C}.
The polarization of the hard component is assumed to originate from the BL/SL. 
The polarization of the soft photons reflected off from the accretion disc also contributes to the observed PD (\citealt{2023MNRAS.519.3681F}). 
In Z-sources, studies suggest that the observed PD is correlated with the position in the Z-track possibly indicating a change in geometry between these states \citep{cocchi2023discovery,Fabiani2024}.
 
The Galactic X-ray source, GX 9+1 (also known as 4U 1758-20 or X Sgr X-3) is a bright, persistent NS-LMXB, classified as an atoll-source located in the direction of Galactic bulge. 
The source is seen to spend most of its time in the soft state (also known as the banana-state) since its discovery in 1965 using the Geiger counters onboard Aerobee rockets by the Naval Research Laboratory \cite{Friedman1967}. 
Later, the location ($l = 9.1^{\circ}$ and $b = 1.2^{\circ}$) was confirmed by \citealt{1968ApJ...152.1005B}. 
The source is located at a distance of 5 kpc as found by \citealt{2005A&A...439..575I} from the estimated value of the equivalent hydrogen column (N$_{H}$) using \textit{BeppoSAX} observations. 
\citealt{2017ApJ...834...71V} identified a NIR counterpart of GX 9+1 after using the revised source position determined using Chandra data.
\citealt{1985SSRv...40..367L} fitted the \textit{EXOSAT} X-ray spectrum of GX 9+1 using a combination of blackbody component and a thermal bremsstrahlung.
Even after a long period since its discovery, many of the system parameters remain hardly constrained. \citealt{2023MNRAS.525.2355T} report an inclination of $29 ^{+3^{\circ}}_{-4}$ for the source, using \textit{AstroSat} and \textit{NuSTAR} observations. 

GX 9+1 was observed by \textit{IXPE} starting from August 31, 2024 and \textit{NICER} on September 1, 2024.
We use the simultaneous \textit{IXPE} and \textit{NICER} observations to perform a spectro-polarimetric study of the source for the first time. 
The paper is organized as follows. 
The observation and data reduction process of \textit{IXPE} and \textit{NICER} data are described in Section \ref{obs}. 
Section \ref{data_ana} deals with the data analysis and presents the results obtained from spectral and spectro-polarimetric analysis. 
We discuss the results and present a summary of the paper, in Sections \ref{discussion}.

\section{Observation and data reduction}\label{obs}

\subsection{IXPE}
\begin{figure}
    \centering
    \includegraphics[width=\linewidth]{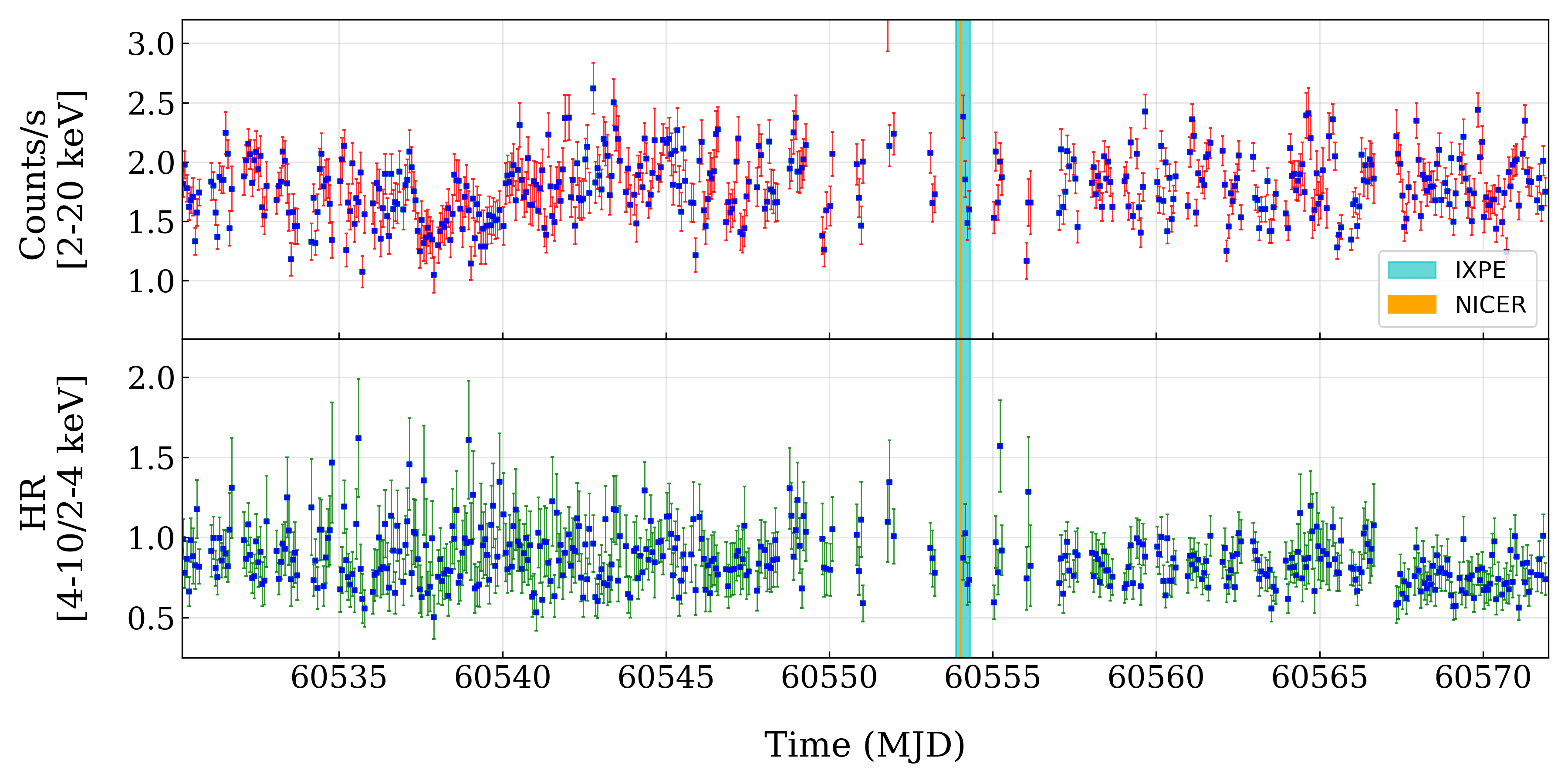}
    \caption{\textit{MAXI}/GSC light curve in the 2–20 keV energy range. The epochs of the \textit{IXPE} and \textit{NICER} observations are marked by cyan and orange color vertical lines respectively.}
    \label{MAXI_lc}
\end{figure}

The Imaging X-ray Polarimetry Explorer (\textit{IXPE}) is capable of providing polarization measurements in the 2-8 keV energy range using three identical telescopes, each consisting of a mirror module assembly and a polarization sensitive imaging detector positioned at the focal plane. 
\textit{IXPE} observed GX 9+1 starting from August 31, 2024 for a net exposure time of approximately 20 ks for each detector units (DUs) 
The details of the observation are provided in Table \ref{obs_log}. 
Figure \ref{MAXI_lc} displays the long-term \textit{MAXI} light curve of the source. 
The figure also shows the period of \textit{IXPE} and \textit{NICER} observations as vertical shaded regions in cyan and orange color respectively, which are simultaneous. 
We note that the source count rate remains stable in the \textit{MAXI} light curve. 
The source hardly shows variability in its long term light curve and remains in the same state throughout the observation period. 

\begin{table*}
    \renewcommand{\arraystretch}{1.2}
    \centering
    \caption{Details of \textit{IXPE} and \textit{NICER} observations. The average count rate during the observation is also reported.}
    \begin{tabular}{lcccr}
    \hline
    \hline
    Satellite & Obs. ID & Start time (UTC) & Exposure (s) & Count rate  \\
    \hline
    \textit{IXPE} & 03003801  &  2024-08-31 21:01:33.252 & 20458 & 56  \\
    \textit{NICER} & 7700010102  & 2024-09-01 00:10:18.00  &  203 & 630  \\
    \hline
    \end{tabular}
    \label{obs_log} 
\end{table*}

We downloaded the \textit{IXPE} Level-2 data from the HEASoft archive\footnote{\href{https://heasarc.gsfc.nasa.gov/cgi-bin/W3Browse/w3browse.pl}{https://heasarc.gsfc.nasa.gov/cgi-bin/W3Browse/w3browse.pl}} and used the \texttt{IXPEOBSSIM} software v31.0.0 (\citealt{2022SoftX..1901194B}) to analyze the data. 
Source counts were extracted from a circular region of 100 arc seconds radius centered around the intensity centroid as recommended by \citealt{DiMarco_2023}. 
We did not perform background subtraction because the source was bright and the background is negligible (\citealt{DiMarco_2023}). 
The source event files are extracted using the \texttt{XPSELECT} task. 
The latest calibration files and response matrices (v13) are used during the analysis.
Various binning algorithms are applied using the \texttt{XPBIN} task. 
We used \texttt{PCUBE} algorithm (\citealt{2015APh....68...45K}) to obtain model-independent polarization parameters such as the polarization degree (PD) and polarization angle (PA). 
We extracted the Stokes I, Q and U spectra and light curve using the \texttt{XPBIN} task.
Spectral analysis was performed using HEASoft\footnote{\href{https://heasarc.gsfc.nasa.gov/docs/software/HEASoft}{https://heasarc.gsfc.nasa.gov/docs/software/HEASoft}} v6.33 and standard ftool tasks.
Figure \ref{IXPE_lc} shows the \textit{IXPE} light curve plotted with a bin size of 150 s in the 2-8 keV energy range.
\subsection{NICER}
The X-ray Timing instrument (XTI) in the Neutron Star Interior Composition Explorer (\textit{NICER}; \citealt{2012SPIE.8443E..13G}) mission offers X-ray data in the 0.2-12.0 keV energy range. 
The XTI consists of 56 X-ray concentrator optics and a silicon drift detector (SDD) positioned at the focal plane. 
\textit{NICER} observed GX 9+1 on September 1, 2024, with an effective exposure of 175 s (see Table \ref{obs_log}), which is simultaneous with the \textit{IXPE} observations.
Despite the low exposure, the high brightness of the source allowed sufficient statistics to be obtained for spectral studies.
Due to the light leakage reported for \textit{NICER} in observations conducted after 22 May 2023, it is recommended to use HEASoft 6.32 or higher for data reduction. 
It automatically selects only nominal-threshold data during data reduction.
We reduced the \textit{NICER} data using the standard data analysis software (\texttt{NICERDAS}, v.2.1.2), which is part of HEASoft v6.33.1, along with the latest CALDB files (v20240206), following the standard procedures outlined in the \textit{NICER} data analysis threads.\footnote{\href{https://heasarc.gsfc.nasa.gov/docs/nicer/analysis_threads/}{https://heasarc.gsfc.nasa.gov/docs/nicer/analysis\_threads/}}.
We generated cleaned event files from unfiltered event files using the \texttt{nicerl2} routine and created the spectrum and light curve using \texttt{nicerl3-spec} and \texttt{nicerl3-lc} routines, respectively. 
The script \texttt{nicerl3-spec} also generates the associated response files for the observation. 
The background files for the observations are simulated using the \texttt{nibackgen3C502} task (\citealt{Remillard_2022}). 
The \textit{NICER} spectra thus generated are grouped along with the response files and background files and rebinned using the \texttt{grppha} task in HEASoft to have a minimum of 25 counts per bin. 
The spectral data in the energy range 0.6-11 keV is considered for spectral fitting. 
We ignored the 11-12 keV energy range due to the lack of statistics in this range.

\section{Data analysis and results}\label{data_ana}
\subsection{Light curve and HID}
\begin{figure}
    \centering
    \includegraphics[width=\linewidth]{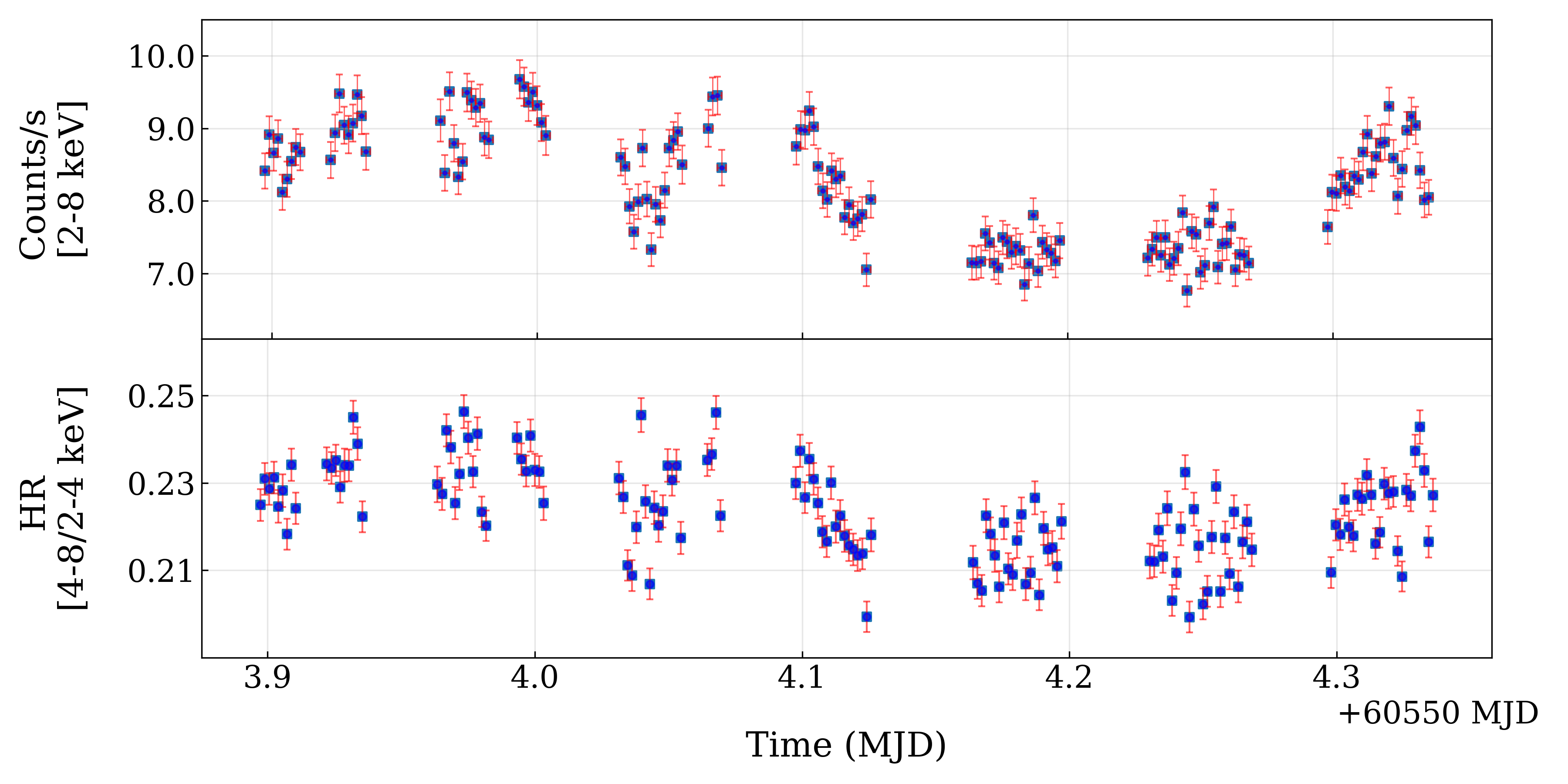}
    \caption{\textit{IXPE} light curve of GX 9+1 (combining the three detector units) plotted for a binsize of 150 s in the 2-8 keV energy band. The start time of the observation is taken as zero. }
    \label{IXPE_lc}
\end{figure}

\begin{figure}
    \centering
    \includegraphics[width=0.9\linewidth]{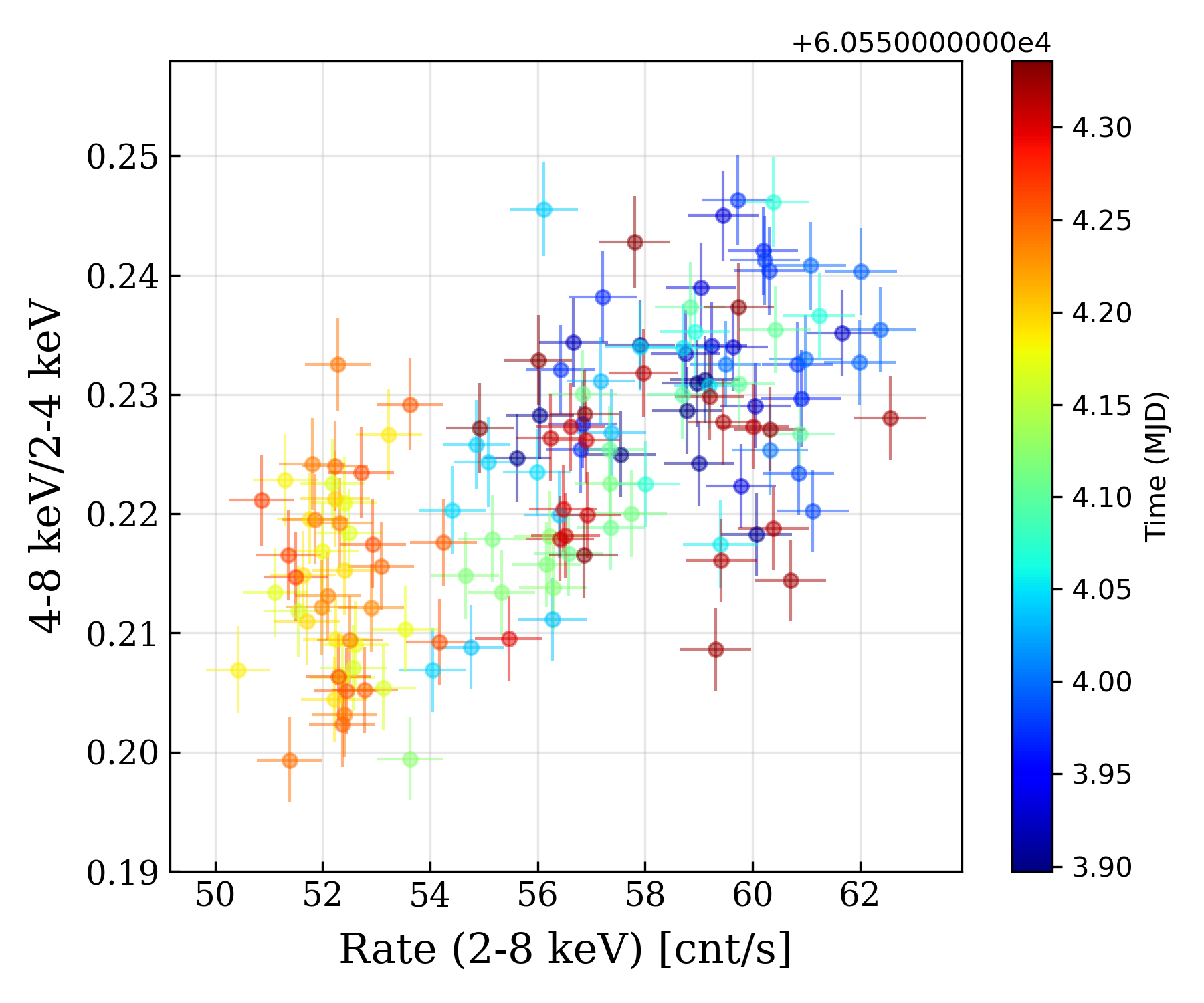}
    \caption{\textit{IXPE} HID plotted with a bin size of 150 s. The color represents the time reference with respect to the start of observation. The source stays in the banana state throughout the observation.}
    \label{IXPE_hid}
\end{figure}

The \textit{IXPE} light curve that combines the three DUs in the energy range of 2-8 keV during observation is shown in Figure \ref{IXPE_lc}. 
The light curve is plotted with a bin size of 150 s, with an X-axis in MJD and Y-axis the combined count rate from the three DUs. 
Significant variations in count rate are observed in the light curve during the observation.
The source count rate shows a drop for a period of $\sim$ 10 ks. 
Hardness-ratio is computed by considering 2-4 keV and 4-8 keV energy bands as the soft and hard bands respectively. 
The variation in hardness ratio is found to be correlated with the variation in the count rate.
Figure \ref{IXPE_hid} shows the HID generated using \textit{IXPE} observation. 
A color-code is given to each data point to show the movement of source along the CCD as a function of time.
No state change is observed and the source remains in the soft state throughout the observation.

\subsection{Spectral properties}
The spectral analysis is performed using \texttt{XSPEC} v.12.14.0. (\citealt{1996ASPC..101...17A}).
We performed a joint fit to the simultaneous \textit{IXPE} and \textit{NICER} spectra in \texttt{XSPEC}.
The \textit{IXPE} observation (refer Figure \ref{IXPE_hid}) shows a lack of state change during the observations. 
We use the data from \textit{NICER} (0.6-11 keV) and from \textit{IXPE} (2-8 keV) for spectral fitting. 
During the fitting, we include a cross-calibration multiplicative constant factor for each DUs of \textit{IXPE} and \textit{NICER} spectra.
The factor is fixed to 1, for \textit{NICER} spectrum and allowed it to vary freely for the three \textit{IXPE} DU spectra.
The \textit{tbabs} model included in the spectral fit takes into account for the interstellar absorption along the line of sight.
The neutral hydrogen column density ($N_{H}$) is allowed to vary during the spectral fit. 

Initially, we attempt to model the continuum in the 0.6-11 keV using a combination of multi-color disc blackbody emission (\textit{diskbb}; \citealt{1984PASJ...36..741M}) from the accretion disc and a thermally Comptonized continuum model (\textit{nthComp}; \citealt{1996MNRAS.283..193Z}).
The spectral fit using this model (\textit{tbabs*(diskbb+nthComp)}) was not satisfactory with a reduced chi-square ($\chi^2/dof$) close to 2. 
Moreover, the fit also resulted in unacceptably high values for the electron temperature ($kT_{e}$) of the corona.
Modification of the model with the inclusion of a \textit{ power law} or a \textit{blackbody} in place of the \textit{nthcomp} component resulted in nonphysical values for the inner disc temperature, and these models are also rejected.

\begin{figure}
    \centering
    \includegraphics[width=0.9\linewidth]{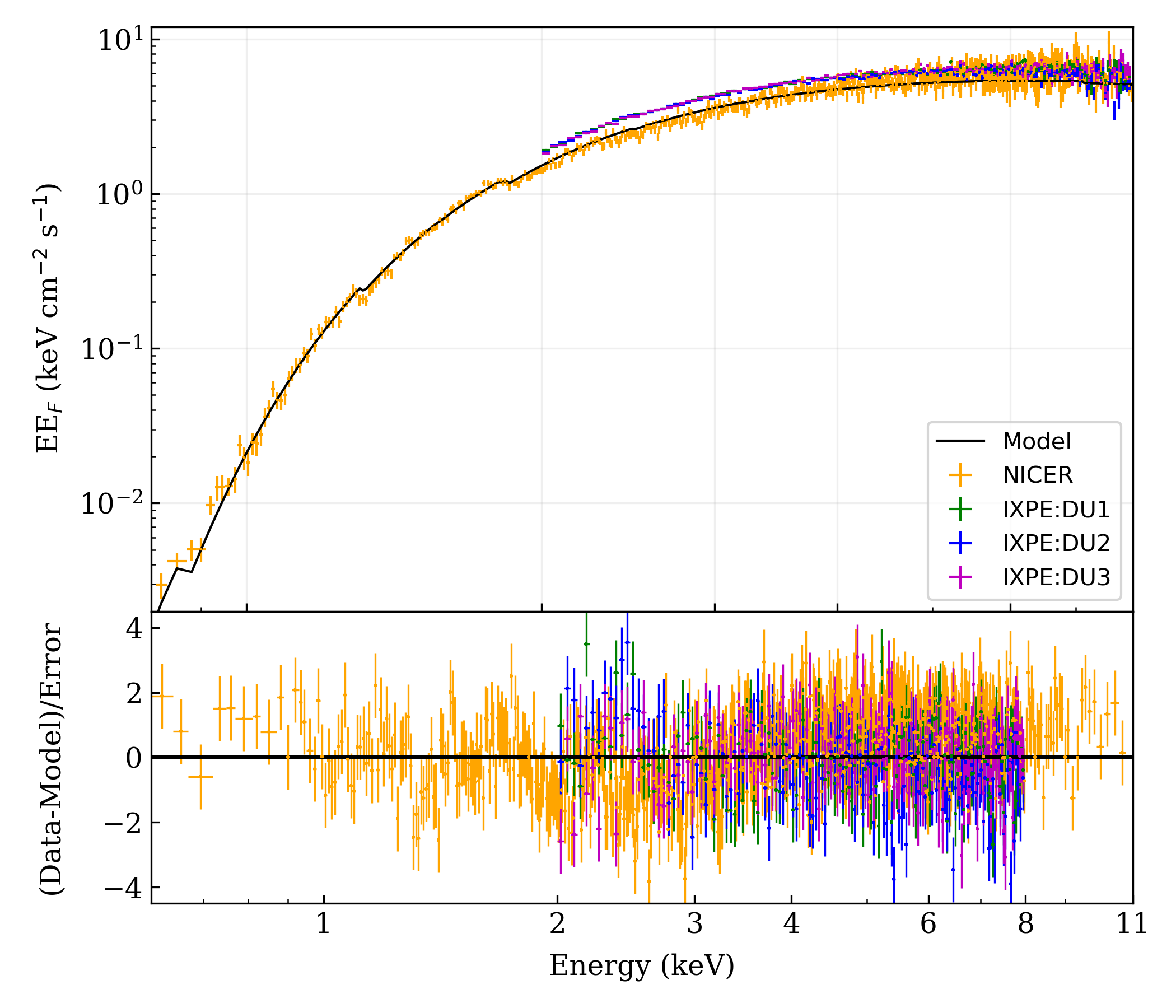}
    \caption{Spectrum of GX 9+1 in EF$_{E}$ representation.
The \textit{NICER} data is plotted in orange color. \textit{IXPE} spectra from three DUs are plotted in cyan, blue and magenta. The model is reported in black lines. The bottom panel shows the residuals between the data and the best-fit model.}
    \label{ixpe_nicer_spec}
\end{figure}

We modeled the simultaneous \textit{NICER} and \textit{IXPE} spectrum using a combination of Comptonized blackbody emission from the neutron star surface and represented by \textit{compbb} (\citealt{1986PASJ...38..819N}) component and a thermal Comptonized continuum with disk-blackbody seed photons, described by the \textit{nthcomp} component.
The model is multiplied with \textit{tbabs} for taking into consideration of the interstellar absorption along the line of sight. 
The model provides a satisfactory fit to the spectrum in the 0.6–11.0 keV energy range, with a value of $\chi^2/\mathrm{dof}$ of 1403/1179.
The final model, which provided the best spectral fit to the combined \textit{NICER} and \textit{IXPE} data is;
\[tbabs*(compbb+nthcomp)\]
The \textit{nthcomp} model includes the photon index ($\Gamma$), electron temperature ($kT_e$) and the blackbody seed photon temperature ($kT_{bb}$) as free parameters.
The \textit{nthcomp} model gives the provision for selecting the type of seed photons. 
We select seed photons to be the disc photons by setting the \textit{inp\_type} parameter to 1 during spectral fitting.
The \textit{compbb} model has electron temperature ($kT_e$), blackbody seed photon temperature ($kT_{bb}$) and optical depth ($\tau$) of the plasma as free parameters.
We initially allowed the $N_{H}$ value to vary freely, but later fixed it to the best fit value of 1.71.
The best-fit spectral parameters for the model are reported in Table \ref{nicer_ixpe_fit_model2}.
The spectral fit to the simultaneous \textit{IXPE} and \textit{NICER} spectra using this model is shown in Figure \ref{ixpe_nicer_spec}.
We estimate the electron temperature to be $\sim$ 2.4 keV and $\sim$ 1.8 respectively, for the \textit{compbb} and \textit{nthcomp} components. 
The photon index ($\Gamma$) for the \textit{nthcomp} is $\sim$1.72.
The optical depth of the plasma was found to be $\sim$ 10 from the \textit{compbb} model.
Since we were unable to constrain the seed photon temperature of the \textit{nthcomp} and \textit{compbb} components, we fixed it at the best-fit values of 0.79 and 0.77 keV respectively.
We also compute the optical depth using the $\Gamma$ value estimated from the \textit{nthcomp} model assuming a spherical geometry for the corona using the relation given by \citealt{1996MNRAS.283..193Z}; 

\begin{equation*}
     \Gamma = -\frac{3}{2}+\sqrt{\frac{9}{4}+\frac{1}{\frac{kT_e}{mc^2}\big(1+\frac{\tau}{3}\big)\tau}}
\end{equation*}

The optical depth is estimated to be 16.3$^{+0.8}_{-0.4}$ and is much higher compared to the one obtained from the \textit{compbb} model.
The best-fit parameters obtained from the spectral fit are comparable to the typical values for atoll-sources in the soft state.

\begin{table}
    \renewcommand{\arraystretch}{1.3}
    \centering
    \caption{The best-fit parameters of simultaneous \textit{NICER} and \textit{IXPE} spectral fitting using the model \textit{tbabs*(compbb*nthcomp)}.}
    \begin{tabular}{lcc}
    \hline
    \hline
     Model & Parameter  & Value  \\
     \hline
    \textit{const} & NICER & 1$^{*}$ \\
	  & IXPE: DU1 & 0.832 $\pm$ 0.005 \\
	  & IXPE: DU2 & 0.841 $\pm$ 0.005 \\
	  & IXPE: DU3 & 0.832 $\pm$ 0.005 \\
      \hline
    \textit{tbabs} & N$_{H}$($\times$10$^{22}$ cm$^{-2}$) & 1.71$^{*}$ \\
    \hline
    \textit{compbb} & $kT_{bb}$ (keV) &  0.79$^{*}$ \\
                & $kT_e$ (keV) &  2.4 $^{+0.4}_{-0.7}$   \\
                & $\tau$ &  10.0 $^{+0.46}_{-0.46}$  \\
                &  N$_{compbb}$ ($\times$ 10$^3$) &  2.1 $^{+0.9}_{-1.6}$  \\
    \hline
    \textit{nthComp} & $kT_{bb}$ (keV) &  0.77$^{*}$ \\
            & $\Gamma$ &  1.72 $^{+0.02}_{-0.05}$  \\
    		& $kT_{e}$ (keV) & 1.81 $^{+0.06}_{-0.06}$ \\
    		& N$_{nthcomp}$ & 1.37 $^{+0.07}_{-0.03}$ \\
            & $\tau_{nthcomp}$ & 16.32 $^{+0.79}_{-0.41}$ \\
    \hline
            & $F_{compbb}$ $ (10^{-9} \, \text{erg} \, \text{s}^{-1} \, \text{cm}^{-2}) $  & 4 $\pm$ 1 \\
            & $F_{nthcomp}$ $(10^{-9} \, \text{erg} \, \text{s}^{-1} \, \text{cm}^{-2})$ & 5 $\pm$ 1 \\
    \hline
    $\chi^{2}$/dof &  & 1403/1179 \\
    \hline
    \hline
    \end{tabular}
    \label{nicer_ixpe_fit_model2} 
    \flushleft
    $^{*}$ represents frozen parameter \\
    Flux values are calculated in the 0.5-10.0 keV energy range.
\end{table}

\subsection{Spectro-polarimetric properties}
\begin{figure*}
    \centering
    \includegraphics[width=0.24\linewidth]{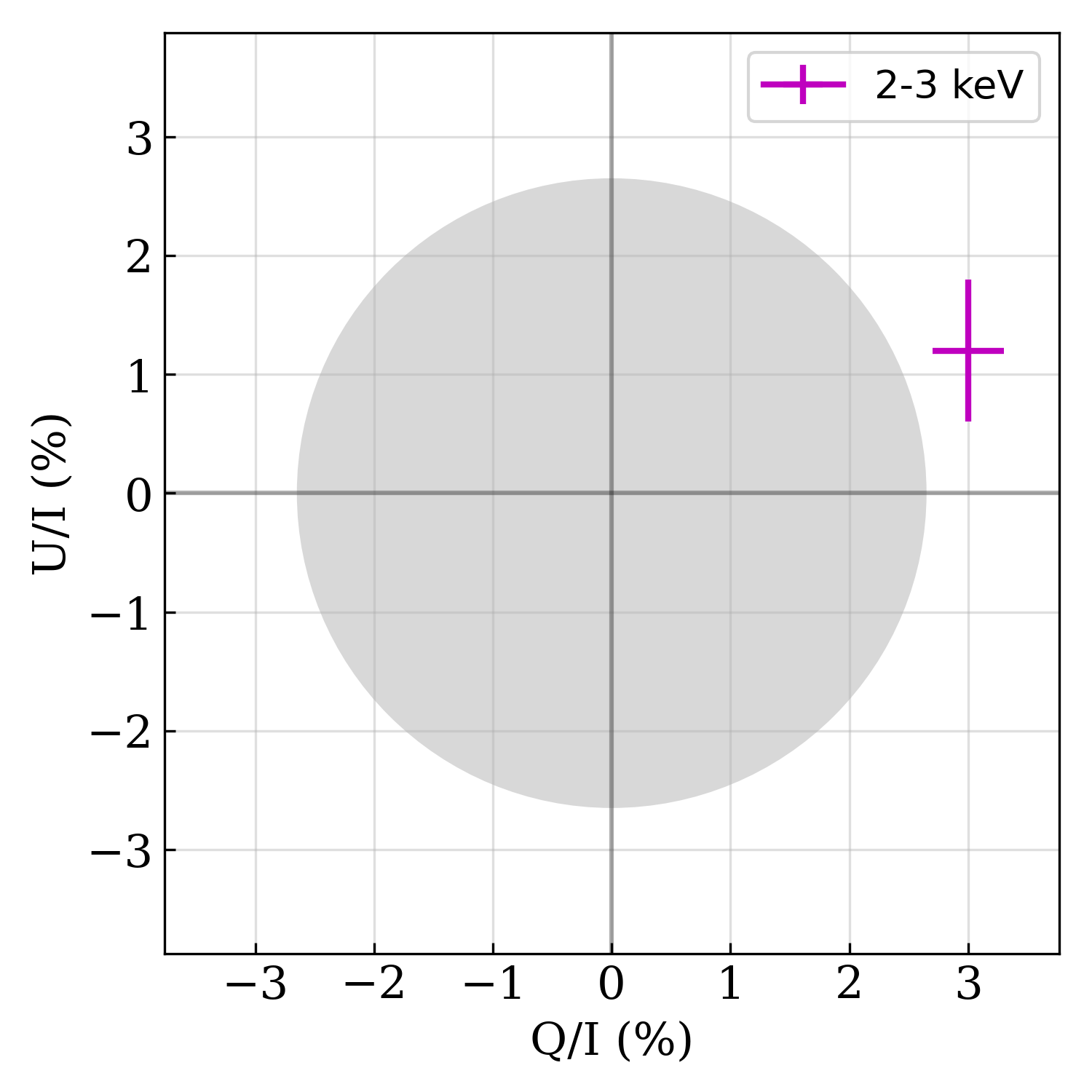}
    \includegraphics[width=0.24\linewidth]{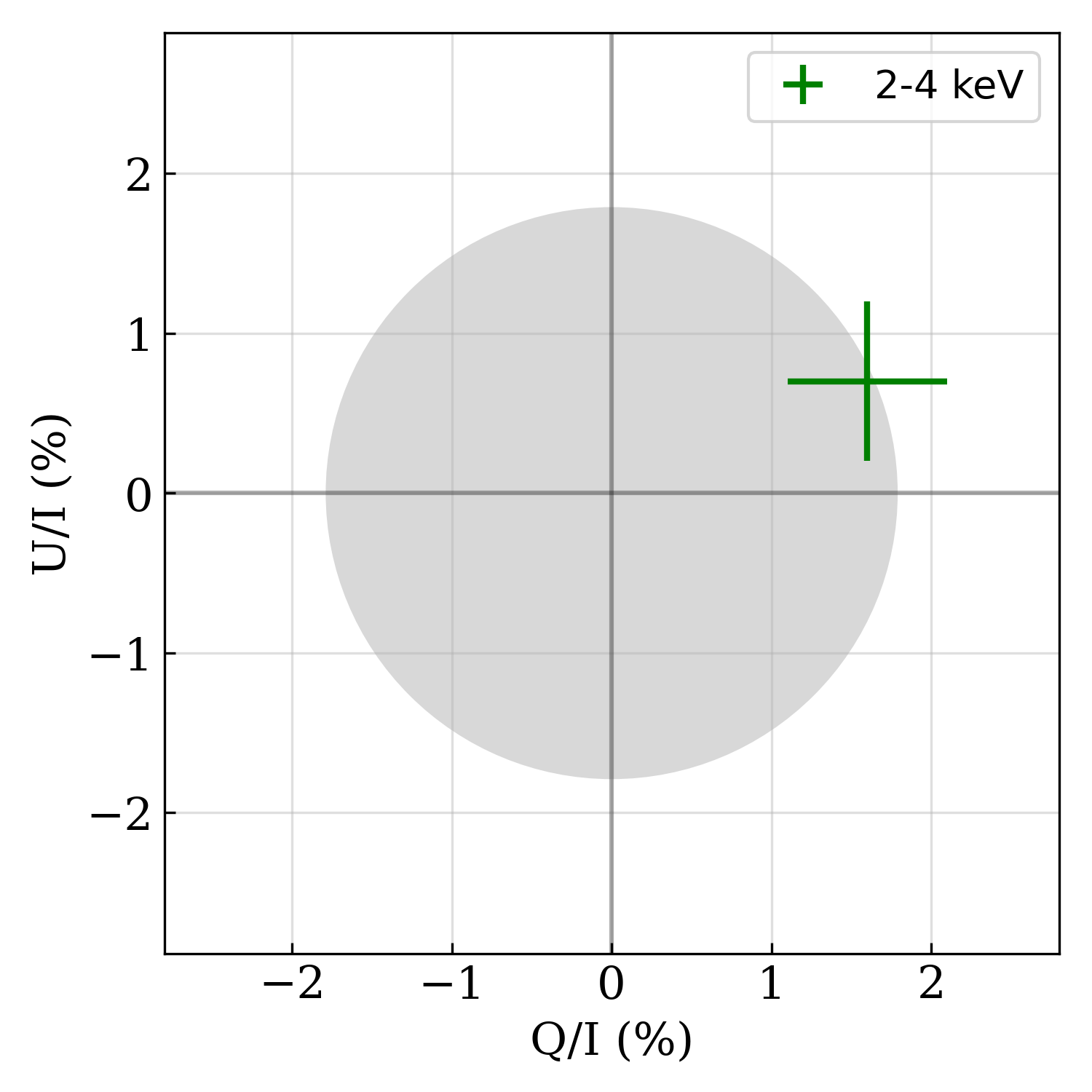}
    \includegraphics[width=0.24\linewidth]{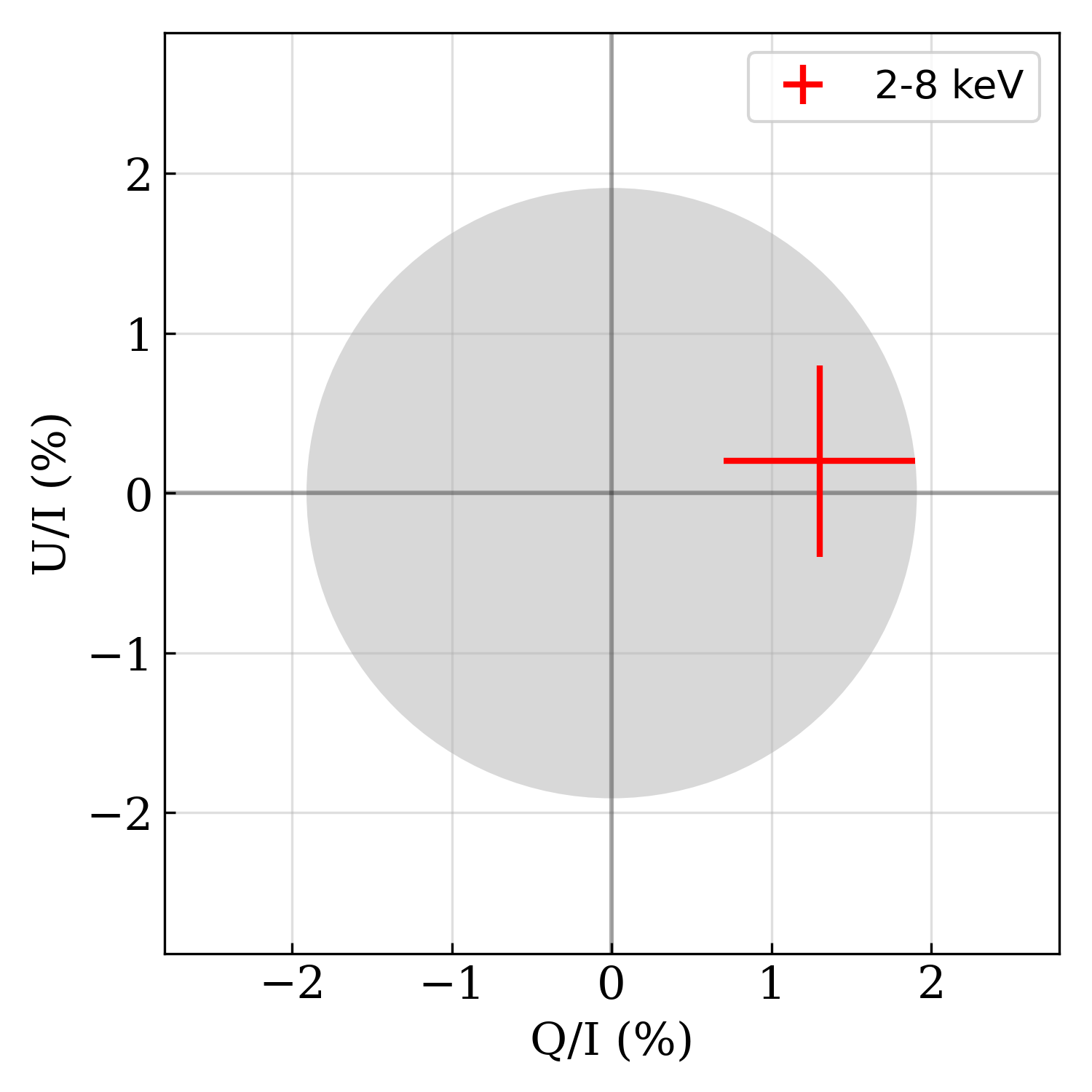}
    \includegraphics[width=0.24\linewidth]{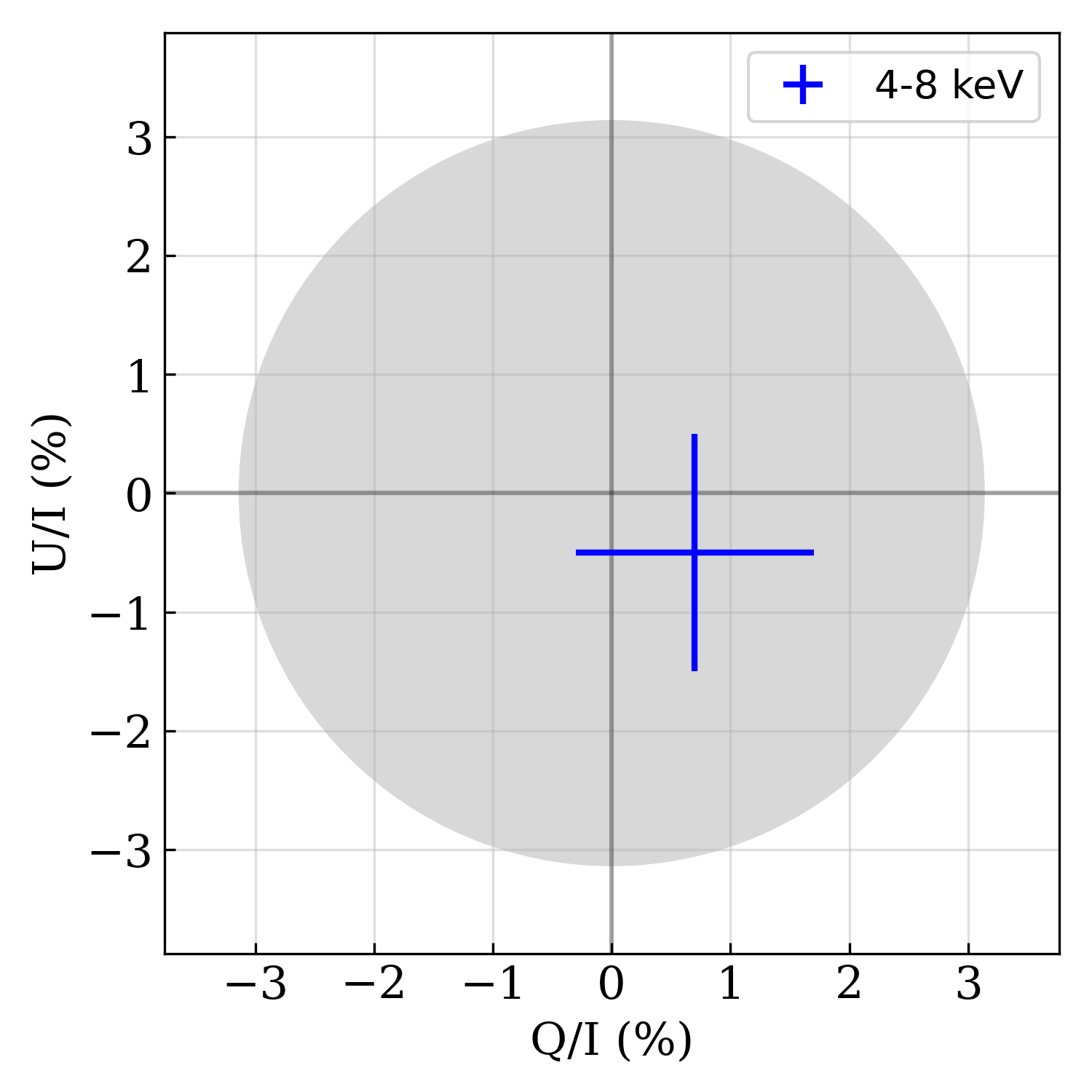}
    \caption{Energy independent dependent polarization behavior derived using \texttt{PCUBE} combining the three DUs. The shaded region denotes the MDP$_{99}$}
    \label{Ene_dep_pd_xpbin}
\end{figure*}

\begin{figure}
    \centering
    \includegraphics[width=0.9\linewidth]{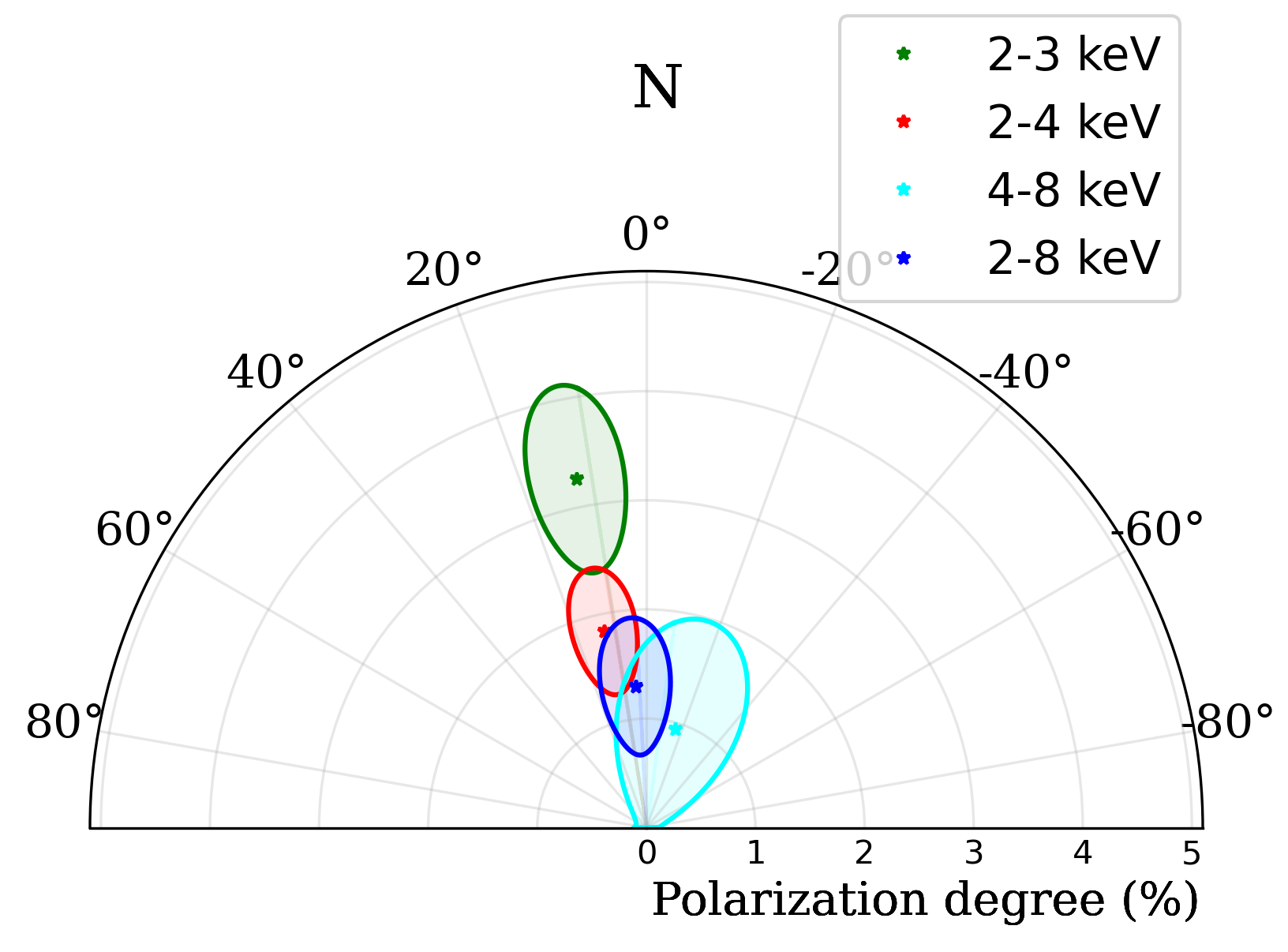}
    \caption{The PD and PA values obtained using the model-independent \texttt{PCUBE} analysis in four energy bands (2-3, 2-4, 4-8 and 2-8 keV). The contour represent 1$\sigma$ confidence interval.}
    \label{pcube_mi}
\end{figure}

To estimate the polarization-degree (PD) and polarization-angle (PA) for the source during the observation, model-independent polarimetric analysis is performed using \texttt{PCUBE} algorithm. 
The latest available calibration file (v.13) is used for the analysis.
The value of PD (1.3 $\pm$ 0.6) obtained from the analysis is found to be less than MDP$_{99}$ in the \textit{IXPE} energy band (2-8 keV).
However, we detect significant polarization ($>$99 \%) in the softer energy range (2-3 keV). 
High PD is observed in the 2-3 keV energy band with a PD 3.3 $\pm$ 0.8\% at an angle PA 11 $\pm$ 7\textdegree.
The results from the model-independent polarimetric analysis is presented in Table \ref{pcube_tab}.
Figure \ref{Ene_dep_pd_xpbin} displays the PD values obtained in four different energy bands (2-3, 2-4, 4-8 and 2-8 keV) and Figure \ref{pcube_mi} shows the 1$\sigma$ contour for the PA and PD values. 
We were unable to derive the energy-dependent polarization properties for the source, as we could not constrain the polarization degree at higher energies.

\begin{table}
    \renewcommand{\arraystretch}{1.3}
    \centering
    \caption{Polarization measured using \texttt{PCUBE} algorithm in different energy bands. The errors are reported in 1$\sigma$ confidence range.}
    \begin{tabular}{lcccc}
    \hline
    \hline
    Parameter & 2-3 keV & 2-4 keV  & 4-8 keV & 2-8 keV  \\
     \hline
    U/I (\%) & 1.2 $\pm$ 0.6 & 0.7 $\pm$ 0.5 & -0.5 $\pm$ 1.0 & 0.2 $\pm$ 0.6 \\
    Q/I (\%) & 3.0 $\pm$ 0.3 & 1.6 $\pm$ 0.5 & 0.7 $\pm$ 1.0 & 1.3 $\pm$ 0.6 \\
    PD (\%) & 3.3 $\pm$ 0.8 & 1.8 $\pm$ 0.6 & 0.9 $\pm$ 1.0 & 1.3 $\pm$ 0.6 \\
    PA (deg) & 11 $\pm$ 7 & 12 $\pm$ 9 & -16 $\pm$ 30 & 4 $\pm$ 13 \\
    MDP$_{99}$ (\%) & 2.65 & 1.79  & 3.14 & 1.91 \\
    \hline
    \end{tabular}
    \label{pcube_tab} 
\end{table} 

We performed the model-dependent polarimetric study by fitting the \textit{IXPE} spectrum for different Stokes parameters (I, U and Q) of all DUs using \textit{polconst} model.
The spectral parameters are fixed to the respective best-fit values reported in Table \ref{nicer_ixpe_fit_model2}.
We allowed the normalization of the \textit{compbb}, and \textit{nthcomp} components to freely vary during the fit.
The PA and PD values derived from the spectral fit in the 2-8 keV \textit{IXPE} energy band is consistent with the values obtained from the model-independent study using the \texttt{PCUBE} algorithm.

To estimate the fraction of polarization for each spectral component, we multiplied \textit{polconst} with each of the components in the model.
Assuming that only one component is polarized, we fix the PD to zero for the other component.
Table \ref{pol_contribution} shows the results from the analysis using the spectral model.
The result suggests that \textit{compbb} has a PD of 5 $\pm$ 4\% at an angle of 3$^{\circ} \pm 37^{\circ}$ at 1.6$\sigma$ (90\%) confidence level. 
We derive an upper-limit to the polarization of Comptonized blackbody component to be 12.2\% at 3$\sigma$ confidence limit.
We also note that the Comptonization continuum component from the disc \textit{nthcomp} is less polarized compared to the blackbody component at 1.2 $\pm$ 0.9\% (90\% confidence limit). 
The upper limit derived for this component is 2. 9\% at that the 3$\sigma$ confidence limit.

\begin{table}
	\renewcommand{\arraystretch}{1.3}
	\centering
	\caption{Polarization of individual components estimated after fixing the PD of other component to zero. The upper limits are reported at 3$\sigma$ confidence level}
	\vspace{0.4cm}
	\label{pol_contribution}
	\begin{tabular}{lccr} 
        \hline
        \hline
	Component & PD (\%) & PA (deg) \\
	\hline
        \multicolumn{3}{c}{3$\sigma$ (99.97\% confidence level)}\\
        \hline
        \textit{compbb} &   $<$ 12.2  &   Unconstrained    \\
        \textit{nthcomp} &   0$^{*}$  &    --   \\
	\hline
        \textit{compbb} &   0$^{*}$   &   --    \\
        \textit{nthcomp} &  $<$ 2.9  &   Unconstrained   \\
        \hline
        \multicolumn{3}{c}{1.6$\sigma$ (90\% confidence level)}\\
        \hline
        \textit{compbb} &   5 $\pm$ 4  &  3 $\pm$ 37    \\
        \textit{nthcomp} &   0$^{*}$  &    --   \\
	\hline
        \textit{compbb} &  0$^{*}$   &   --    \\
        \textit{nthcomp} &  1.2 $\pm$ 0.9  &  4 $\pm$ 24   \\
        \hline
        \hline
	\end{tabular}
    \flushleft
    $^{*}$ represents frozen parameter
\end{table}

To understand the energy-dependent behavior of polarization parameters, we attempted by splitting the \textit{IXPE} energy band into smaller energy bins. But there was no significant detection in the energy bands above 4 keV. 
Splitting the data in time also resulted in no significant detection of polarization in the segments.

\section{Discussion}\label{discussion}
This paper discusses the results from the first spectro-polarimetric observation of atoll NS-LMXB GX 9+1. 
We performed a joint spectral fit using simultaneous \textit{IXPE}, and \textit{NICER} observations. 
The source is observed to be in the soft state throughout the observation.
The combined \textit{NICER} and \textit{IXPE} X-ray spectrum could be well modeled using a combination of Comptonized blackbody emission from the neutron star surface and Comptonized continuum with disc photons as input seed photons. 
No iron line or other reflection features are observed in the 0.6-11 keV spectrum.
The observed PD in the 2-8 keV energy range is below the minimum detectable polarization (MDP$_{99}$).
However, a PD of 1.3\% at 1.55$\sigma$ (88\% confidence level) is detected in the 2-8 keV energy band.
A significant polarization with a PD of $\sim$3\% (at the 3.3$\sigma$ confidence level) is detected at a PA of $\sim$12$^{\circ}$ in the 2-3 keV energy range.
The low polarization of 1.3\% in the \textit{IXPE} energy range matches with the observed PD and the determined upper limit for PD in atoll-sources in soft state studied so far using \textit{IXPE} \citep{2023ApJ...943..129C, 2023A&A...676A..20U, 2023ApJ...953L..22D, 2024A&A...692A.123G}.
However, in sources which shows the presence of reflection, a higher value of PD is observed (PD of $\sim$1.7\% for GX 9+9 (\citealt{2023MNRAS.521L..74C}). 
No evidence for the presence of reflection features have been observed in GX 9+1.
In all atoll-sources observed so far using \textit{IXPE}, a low polarization is found in the 2–4 keV energy band, where the thermal emission dominates, compared to the 4-8 keV energy band. 
The observed PD of 1.8\% in the 2-4 keV and 3\% in 2-3 keV in GX 9+1 are contradictory to these earlier observations.
No significant polarization is detected in the 4-8 keV range.
We derive upper-limits to the degree of polarization of the Comptonized blackbody component as well as the Comptonized continuum component and note that the comptonized blackbody component is more polarized. 
The observed high PD in the soft energy band is attributed to the polarization of the 
Comptonized blackbody component in the spectrum.

\begin{figure}
    \centering
    \includegraphics[width=0.9\linewidth]{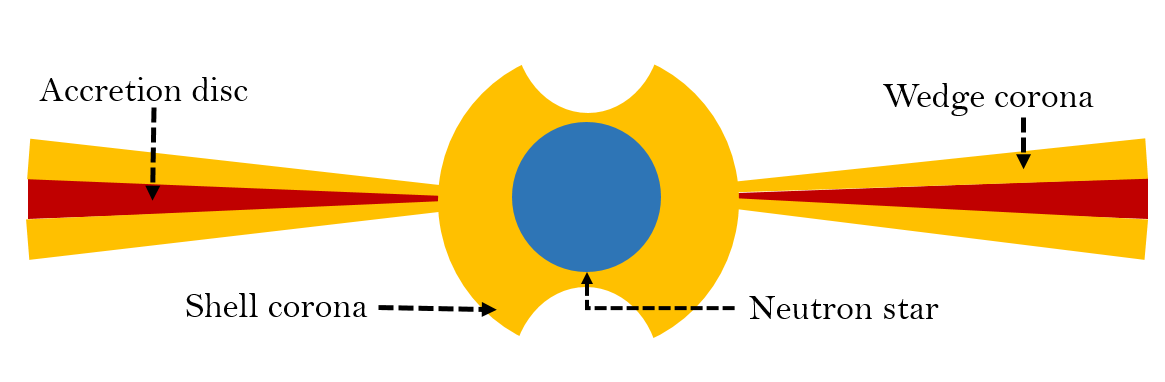}
    \caption{Schematic representation of the coronal geometries inferred from the polarimetric properties. A sandwich and shell geometry with polar caps removed are proposed for the source.}
    \label{geometry}
\end{figure}

The observed PD in the source depends both on the geometry of the Comptonizing region and the inclination of the source to the observer's line of sight.
The source is believed to have a low inclination of $\sim$30$^{\circ}$, as estimated by \citealt{2023MNRAS.525.2355T}.
Spectral modeling suggests that a combination of Comptonized photons from the accretion disc as well as the neutron star surface, well describes the observed X-ray spectrum.
The emission from the accretion disc is not directly observed.
We assume an optically thick, wedge-shaped corona located above the accretion disc gives rise to Comptonized emission from the disc (\textit{nthcomp} component) in the spectrum.
Thermal seed photons from the disc are scattered off by hot electrons in the corona, leading to the emission at high energies.
The nature of polarization properties in a similar scenario, were the disc and corona are in sandwich geometry, was studied in the case of stellar-mass black holes by \citealt{Schnittman_2010}.
Even though the study has been conducted for black hole sources, the results obtained for high luminosity states may be used for an informative comparison, since GX 9+1 was observed in the soft state.
In the soft state, the accretion disc extends close to the compact object similar to what we expect for a black hole source in the same state.
The study suggest that at higher luminosity, a low degree of polarization is expected from a low-inclination system in the \textit{IXPE} energy band (see Figure 3:\citealt{Schnittman_2010}) for a sandwich corona geometry.

The presence of a transition shell (\citealt{Titarchuk_1998}) between the inner accretion disc and the NS surface intercept the blackbody emission from the NS surface leading to a Comptonized blackbody emission in the X-ray spectrum. 
This emission can give rise to the observed polarization signatures in the source.
The detailed simulations by \citealt{2022MNRAS.514.2561G} gives a comparison between the polarization properties of a slab and shell geometries of corona. 
This study suggests that there is a high degree of polarization at lower energies, and that the degree of polarization decreases with energy for a stationary, non-rotating corona with a shell geometry.
This explains the observed PD values of approximately 3\% and 1.8\% in the 2-3 keV and 2-4 keV energy bands, respectively, as well as the null detection of polarization in the 4-8 keV band.
However, on comparison of the PD value, the observed value is expected for a source with high inclination in the shell geometry. 
A shell geometry with top and bottom spherical caps removed can break the symmetry and can help in explaining the high PD observed.
This geometry is similar to the one used by \citealt{2022ApJ...924L..13L} to describe the polarization signatures observed in Sco X-1.
The schematic representations of the coronal geometry is displayed in Figure \ref{geometry}.
In conclusion, the polarization signatures in GX 9+1 favor the presence of an optically thick wedge shaped corona above the accretion disc and a transition shell with polar caps removed, close to the NS surface.
Further observation of longer durations can provide more details regarding the energy-dependent polarization behavior in the source.

\section*{Acknowledgment}
This work is based on observations performed using \textit{IXPE} (MSFC, SSDC, and INFN), and \textit{NuSTAR} satellites. Authors thank GH , SAG; DD, PDMSA and Director, URSC and Department of Physics, University of Calicut for encouragement and support to carry out this research. We have used data software provided by the High Energy Astrophysics Science Archive Research Center (HEASARC), which is a service of the Astrophysics Science Division at NASA/GSFC and the High Energy Astrophysics Division of the Smithsonian Astrophysical Observatory.

\section*{Data availability}
The \textit{NICER} and \textit{IXPE} data used in this work are available at High Energy Astrophysics Science Archive Research Center (\href{https://heasarc.gsfc.nasa.gov/cgi-bin/W3Browse/w3browse.pl}{HEASARC}) facility, located at NASA-Goddard Space Flight Center. 
The \textit{MAXI} light curve used in this work is publicly available at \href{http://maxi.riken.jp/top/slist.html}{http://maxi.riken.jp/top/slist.html}


\bibliographystyle{mnras}
\bibliography{example} 

\begin{thebibliography}{}
\makeatletter
\relax
\def\mn@urlcharsother{\let\do\@makeother \do\$\do\&\do\#\do\^\do\_\do\%\do\~}
\def\mn@doi{\begingroup\mn@urlcharsother \@ifnextchar [ {\mn@doi@} {\mn@doi@[]}}
\def\mn@doi@[#1]#2{\def\@tempa{#1}\ifx\@tempa\@empty \href {http://dx.doi.org/#2} {doi:#2}\else \href {http://dx.doi.org/#2} {#1}\fi \endgroup}
\def\mn@eprint#1#2{\mn@eprint@#1:#2::\@nil}
\def\mn@eprint@arXiv#1{\href {http://arxiv.org/abs/#1} {{\tt arXiv:#1}}}
\def\mn@eprint@dblp#1{\href {http://dblp.uni-trier.de/rec/bibtex/#1.xml} {dblp:#1}}
\def\mn@eprint@#1:#2:#3:#4\@nil{\def\@tempa {#1}\def\@tempb {#2}\def\@tempc {#3}\ifx \@tempc \@empty \let \@tempc \@tempb \let \@tempb \@tempa \fi \ifx \@tempb \@empty \def\@tempb {arXiv}\fi \@ifundefined {mn@eprint@\@tempb}{\@tempb:\@tempc}{\expandafter \expandafter \csname mn@eprint@\@tempb\endcsname \expandafter{\@tempc}}}

\bibitem[\protect\citeauthoryear{Agrawal \& Sreekumar}{Agrawal \& Sreekumar}{2003}]{10.1111/j.1365-2966.2003.07147.x}
Agrawal V.~K.,  Sreekumar P.,  2003, \mn@doi [Monthly Notices of the Royal Astronomical Society] {10.1111/j.1365-2966.2003.07147.x}, 346, 933

\bibitem[\protect\citeauthoryear{Agrawal, Nandi  \& Katoch}{Agrawal et~al.}{2022}]{Agrawal:2022jbm}
Agrawal V.~K.,  Nandi A.,   Katoch T.,  2022, \mn@doi [Mon. Not. Roy. Astron. Soc.] {10.1093/mnras/stac2579}, 518, 194

\bibitem[\protect\citeauthoryear{{Altamirano} et~al.,}{{Altamirano} et~al.}{2010}]{2010ATel.2952....1A}
{Altamirano} D.,  et~al., 2010, The Astronomer's Telegram, \href {https://ui.adsabs.harvard.edu/abs/2010ATel.2952....1A} {2952, 1}

\bibitem[\protect\citeauthoryear{{Arnaud}}{{Arnaud}}{1996}]{1996ASPC..101...17A}
{Arnaud} K.~A.,  1996, in {Jacoby} G.~H.,  {Barnes} J.,  eds,  Astronomical Society of the Pacific Conference Series Vol. 101, Astronomical Data Analysis Software and Systems V. p.~17

\bibitem[\protect\citeauthoryear{{Baldini} et~al.,}{{Baldini} et~al.}{2022}]{2022SoftX..1901194B}
{Baldini} L.,  et~al., 2022, \mn@doi [SoftwareX] {10.1016/j.softx.2022.101194}, \href {https://ui.adsabs.harvard.edu/abs/2022SoftX..1901194B} {19, 101194}

\bibitem[\protect\citeauthoryear{{Barret} \& {Olive}}{{Barret} \& {Olive}}{2002}]{2002ApJ...576..391B}
{Barret} D.,  {Olive} J.-F.,  2002, \mn@doi [apj] {10.1086/341626}, \href {https://ui.adsabs.harvard.edu/abs/2002ApJ...576..391B} {576, 391}

\bibitem[\protect\citeauthoryear{Barret, Olive, Boirin, Done, Skinner  \& Grindlay}{Barret et~al.}{2000}]{Barret_2000}
Barret D.,  Olive J.~F.,  Boirin L.,  Done C.,  Skinner G.~K.,   Grindlay J.~E.,  2000, \mn@doi [The Astrophysical Journal] {10.1086/308651}, 533, 329

\bibitem[\protect\citeauthoryear{{Bradt}, {Naranan}, {Rappaport}  \& {Spada}}{{Bradt} et~al.}{1968}]{1968ApJ...152.1005B}
{Bradt} H.,  {Naranan} S.,  {Rappaport} S.,   {Spada} G.,  1968, \mn@doi [apj] {10.1086/149613}, \href {https://ui.adsabs.harvard.edu/abs/1968ApJ...152.1005B} {152, 1005}

\bibitem[\protect\citeauthoryear{{Cackett} et~al.,}{{Cackett} et~al.}{2008}]{2008ApJ...674..415C}
{Cackett} E.~M.,  et~al., 2008, \mn@doi [apj] {10.1086/524936}, \href {https://ui.adsabs.harvard.edu/abs/2008ApJ...674..415C} {674, 415}

\bibitem[\protect\citeauthoryear{{Capitanio} et~al.,}{{Capitanio} et~al.}{2023a}]{2023ApJ...943..129C}
{Capitanio} F.,  et~al., 2023a, \mn@doi [apj] {10.3847/1538-4357/acae88}, \href {https://ui.adsabs.harvard.edu/abs/2023ApJ...943..129C} {943, 129}

\bibitem[\protect\citeauthoryear{Capitanio et~al.,}{Capitanio et~al.}{2023b}]{Capitanio_2023}
Capitanio F.,  et~al., 2023b, \mn@doi [The Astrophysical Journal] {10.3847/1538-4357/acae88}, 943, 129

\bibitem[\protect\citeauthoryear{{Chatterjee}, {Agrawal}, {Jayasurya}  \& {Katoch}}{{Chatterjee} et~al.}{2023}]{2023MNRAS.521L..74C}
{Chatterjee} R.,  {Agrawal} V.~K.,  {Jayasurya} K.~M.,   {Katoch} T.,  2023, \mn@doi [\mnras] {10.1093/mnrasl/slad026}, \href {https://ui.adsabs.harvard.edu/abs/2023MNRAS.521L..74C} {521, L74}

\bibitem[\protect\citeauthoryear{Cocchi et~al.,}{Cocchi et~al.}{2023b}]{cocchi2023discovery}
Cocchi M.,  et~al., 2023b, Astronomy \& Astrophysics, 674, L10

\bibitem[\protect\citeauthoryear{{Cocchi} et~al.,}{{Cocchi} et~al.}{2023a}]{2023A&A...674L..10C}
{Cocchi} M.,  et~al., 2023a, \mn@doi [aap] {10.1051/0004-6361/202346275}, \href {https://ui.adsabs.harvard.edu/abs/2023A&A...674L..10C} {674, L10}

\bibitem[\protect\citeauthoryear{{Di Marco} et~al.,}{{Di Marco} et~al.}{2023}]{2023ApJ...953L..22D}
{Di Marco} A.,  et~al., 2023, \mn@doi [apjl] {10.3847/2041-8213/acec6e}, \href {https://ui.adsabs.harvard.edu/abs/2023ApJ...953L..22D} {953, L22}

\bibitem[\protect\citeauthoryear{{Di Salvo} et~al.,}{{Di Salvo} et~al.}{2009}]{2009MNRAS.398.2022D}
{Di Salvo} T.,  et~al., 2009, \mn@doi [mnras] {10.1111/j.1365-2966.2009.15240.x}, \href {https://ui.adsabs.harvard.edu/abs/2009MNRAS.398.2022D} {398, 2022}

\bibitem[\protect\citeauthoryear{{D{\'\i}az Trigo}, {Sidoli}, {Boirin}  \& {Parmar}}{{D{\'\i}az Trigo} et~al.}{2012}]{2012A&A...543A..50D}
{D{\'\i}az Trigo} M.,  {Sidoli} L.,  {Boirin} L.,   {Parmar} A.~N.,  2012, \mn@doi [aap] {10.1051/0004-6361/201219049}, \href {https://ui.adsabs.harvard.edu/abs/2012A&A...543A..50D} {543, A50}

\bibitem[\protect\citeauthoryear{{Fabiani, Sergio} et~al.,}{{Fabiani, Sergio} et~al.}{2024}]{Fabiani2024}
{Fabiani, Sergio} et~al., 2024, \mn@doi [A&A] {10.1051/0004-6361/202347374}, 684, A137

\bibitem[\protect\citeauthoryear{{Fabiani} et~al.,}{{Fabiani} et~al.}{2024}]{2024A&A...684A.137F}
{Fabiani} S.,  et~al., 2024, \mn@doi [aap] {10.1051/0004-6361/202347374}, \href {https://ui.adsabs.harvard.edu/abs/2024A&A...684A.137F} {684, A137}

\bibitem[\protect\citeauthoryear{{Farinelli} et~al.,}{{Farinelli} et~al.}{2023}]{2023MNRAS.519.3681F}
{Farinelli} R.,  et~al., 2023, \mn@doi [mnras] {10.1093/mnras/stac3726}, \href {https://ui.adsabs.harvard.edu/abs/2023MNRAS.519.3681F} {519, 3681}

\bibitem[\protect\citeauthoryear{Friedman, Byram  \& Chubb}{Friedman et~al.}{1967}]{Friedman1967}
Friedman H.,  Byram E.~T.,   Chubb T.~A.,  1967, \mn@doi [Science] {10.1126/science.156.3773.374}, 156, 374

\bibitem[\protect\citeauthoryear{{Gendreau}, {Arzoumanian}  \& {Okajima}}{{Gendreau} et~al.}{2012}]{2012SPIE.8443E..13G}
{Gendreau} K.~C.,  {Arzoumanian} Z.,   {Okajima} T.,  2012, in {Takahashi} T.,  {Murray} S.~S.,   {den Herder} J.-W.~A.,  eds,  Society of Photo-Optical Instrumentation Engineers (SPIE) Conference Series Vol. 8443, Space Telescopes and Instrumentation 2012: Ultraviolet to Gamma Ray. p. 844313, \mn@doi{10.1117/12.926396}

\bibitem[\protect\citeauthoryear{{Gnarini}, {Ursini}, {Matt}, {Bianchi}, {Capitanio}, {Cocchi}, {Farinelli}  \& {Zhang}}{{Gnarini} et~al.}{2022}]{2022MNRAS.514.2561G}
{Gnarini} A.,  {Ursini} F.,  {Matt} G.,  {Bianchi} S.,  {Capitanio} F.,  {Cocchi} M.,  {Farinelli} R.,   {Zhang} W.,  2022, \mn@doi [mnras] {10.1093/mnras/stac1523}, \href {https://ui.adsabs.harvard.edu/abs/2022MNRAS.514.2561G} {514, 2561}

\bibitem[\protect\citeauthoryear{{Gnarini} et~al.,}{{Gnarini} et~al.}{2024a}]{2024A&A...690A.230G}
{Gnarini} A.,  et~al., 2024a, \mn@doi [\aap] {10.1051/0004-6361/202450716}, \href {https://ui.adsabs.harvard.edu/abs/2024A&A...690A.230G} {690, A230}

\bibitem[\protect\citeauthoryear{{Gnarini} et~al.,}{{Gnarini} et~al.}{2024b}]{2024A&A...692A.123G}
{Gnarini} A.,  et~al., 2024b, \mn@doi [aap] {10.1051/0004-6361/202452642}, \href {https://ui.adsabs.harvard.edu/abs/2024A&A...692A.123G} {692, A123}

\bibitem[\protect\citeauthoryear{{Hasinger} \& {van der Klis}}{{Hasinger} \& {van der Klis}}{1989}]{1989A&A...225...79H}
{Hasinger} G.,  {van der Klis} M.,  1989, \aap, \href {https://ui.adsabs.harvard.edu/abs/1989A&A...225...79H} {225, 79}

\bibitem[\protect\citeauthoryear{Homan et~al.,}{Homan et~al.}{2010}]{Homan_2010}
Homan J.,  et~al., 2010, \mn@doi [The Astrophysical Journal] {10.1088/0004-637X/719/1/201}, 719, 201

\bibitem[\protect\citeauthoryear{{Iaria}, {di Salvo}, {Robba}, {Lavagetto}, {Burderi}, {Stella}  \& {van der Klis}}{{Iaria} et~al.}{2005}]{2005A&A...439..575I}
{Iaria} R.,  {di Salvo} T.,  {Robba} N.~R.,  {Lavagetto} G.,  {Burderi} L.,  {Stella} L.,   {van der Klis} M.,  2005, \mn@doi [aap] {10.1051/0004-6361:20042231}, \href {https://ui.adsabs.harvard.edu/abs/2005A&A...439..575I} {439, 575}

\bibitem[\protect\citeauthoryear{Jayasurya, Agrawal  \& Chatterjee}{Jayasurya et~al.}{2023}]{Jayasurya:2023udz}
Jayasurya K.~M.,  Agrawal V.~K.,   Chatterjee R.,  2023, \mn@doi [Mon. Not. Roy. Astron. Soc.] {10.1093/mnras/stad2601}, 525, 4657

\bibitem[\protect\citeauthoryear{Kashyap, Chakraborty, Bhattacharyya  \& Ram}{Kashyap et~al.}{2023}]{10.1093/mnras/stad1606}
Kashyap U.,  Chakraborty M.,  Bhattacharyya S.,   Ram B.,  2023, \mn@doi [Monthly Notices of the Royal Astronomical Society] {10.1093/mnras/stad1606}, 523, 2788

\bibitem[\protect\citeauthoryear{{Kislat}, {Clark}, {Beilicke}  \& {Krawczynski}}{{Kislat} et~al.}{2015}]{2015APh....68...45K}
{Kislat} F.,  {Clark} B.,  {Beilicke} M.,   {Krawczynski} H.,  2015, \mn@doi [Astroparticle Physics] {10.1016/j.astropartphys.2015.02.007}, \href {https://ui.adsabs.harvard.edu/abs/2015APh....68...45K} {68, 45}

\bibitem[\protect\citeauthoryear{{Kuulkers}, {van der Klis}, {Oosterbroek}, {Asai}, {Dotani}, {van Paradijs}  \& {Lewin}}{{Kuulkers} et~al.}{1994}]{1994A&A...289..795K}
{Kuulkers} E.,  {van der Klis} M.,  {Oosterbroek} T.,  {Asai} K.,  {Dotani} T.,  {van Paradijs} J.,   {Lewin} W.~H.~G.,  1994, aap, \href {https://ui.adsabs.harvard.edu/abs/1994A&A...289..795K} {289, 795}

\bibitem[\protect\citeauthoryear{{Langmeier}, {Sztajno}, {Truemper}  \& {Hasinger}}{{Langmeier} et~al.}{1985}]{1985SSRv...40..367L}
{Langmeier} A.,  {Sztajno} M.,  {Truemper} J.,   {Hasinger} G.,  1985, \mn@doi [\ssr] {10.1007/BF00179842}, \href {https://ui.adsabs.harvard.edu/abs/1985SSRv...40..367L} {40, 367}

\bibitem[\protect\citeauthoryear{{Lin}, {Remillard}  \& {Homan}}{{Lin} et~al.}{2009}]{2009ApJ...696.1257L}
{Lin} D.,  {Remillard} R.~A.,   {Homan} J.,  2009, \mn@doi [apj] {10.1088/0004-637X/696/2/1257}, \href {https://ui.adsabs.harvard.edu/abs/2009ApJ...696.1257L} {696, 1257}

\bibitem[\protect\citeauthoryear{Lin, Remillard  \& Homan}{Lin et~al.}{2010}]{Lin_2010}
Lin D.,  Remillard R.~A.,   Homan J.,  2010, \mn@doi [The Astrophysical Journal] {10.1088/0004-637X/719/2/1350}, 719, 1350

\bibitem[\protect\citeauthoryear{{Long} et~al.,}{{Long} et~al.}{2022}]{2022ApJ...924L..13L}
{Long} X.,  et~al., 2022, \mn@doi [apjl] {10.3847/2041-8213/ac4673}, \href {https://ui.adsabs.harvard.edu/abs/2022ApJ...924L..13L} {924, L13}

\bibitem[\protect\citeauthoryear{Ludlam et~al.,}{Ludlam et~al.}{2017}]{Ludlam_2017}
Ludlam R.~M.,  et~al., 2017, \mn@doi [The Astrophysical Journal] {10.3847/1538-4357/836/1/140}, 836, 140

\bibitem[\protect\citeauthoryear{Marco et~al.,}{Marco et~al.}{2023}]{DiMarco_2023}
Marco A.~D.,  et~al., 2023, \mn@doi [The Astronomical Journal] {10.3847/1538-3881/acba0f}, 165, 143

\bibitem[\protect\citeauthoryear{{Mitsuda} et~al.,}{{Mitsuda} et~al.}{1984}]{1984PASJ...36..741M}
{Mitsuda} K.,  et~al., 1984, pasj, \href {https://ui.adsabs.harvard.edu/abs/1984PASJ...36..741M} {36, 741}

\bibitem[\protect\citeauthoryear{{Mondal}, {Pahari}, {Dewangan}, {Misra}  \& {Raychaudhuri}}{{Mondal} et~al.}{2017}]{2017MNRAS.466.4991M}
{Mondal} A.~S.,  {Pahari} M.,  {Dewangan} G.~C.,  {Misra} R.,   {Raychaudhuri} B.,  2017, \mn@doi [\mnras] {10.1093/mnras/stx039}, \href {https://ui.adsabs.harvard.edu/abs/2017MNRAS.466.4991M} {466, 4991}

\bibitem[\protect\citeauthoryear{{Nishimura}, {Mitsuda}  \& {Itoh}}{{Nishimura} et~al.}{1986}]{1986PASJ...38..819N}
{Nishimura} J.,  {Mitsuda} K.,   {Itoh} M.,  1986, pasj, \href {https://ui.adsabs.harvard.edu/abs/1986PASJ...38..819N} {38, 819}

\bibitem[\protect\citeauthoryear{{Popham} \& {Sunyaev}}{{Popham} \& {Sunyaev}}{2001}]{2001ApJ...547..355P}
{Popham} R.,  {Sunyaev} R.,  2001, \mn@doi [apj] {10.1086/318336}, \href {https://ui.adsabs.harvard.edu/abs/2001ApJ...547..355P} {547, 355}

\bibitem[\protect\citeauthoryear{Remillard et~al.,}{Remillard et~al.}{2022}]{Remillard_2022}
Remillard R.~A.,  et~al., 2022, \mn@doi [The Astronomical Journal] {10.3847/1538-3881/ac4ae6}, 163, 130

\bibitem[\protect\citeauthoryear{Saade et~al.,}{Saade et~al.}{2024}]{Saade_2024}
Saade M.~L.,  et~al., 2024, \mn@doi [The Astrophysical Journal] {10.3847/1538-4357/ad235a}, 963, 133

\bibitem[\protect\citeauthoryear{Schnittman \& Krolik}{Schnittman \& Krolik}{2010}]{Schnittman_2010}
Schnittman J.~D.,  Krolik J.~H.,  2010, \mn@doi [The Astrophysical Journal] {10.1088/0004-637X/712/2/908}, 712, 908

\bibitem[\protect\citeauthoryear{{Tauris} \& {van den Heuvel}}{{Tauris} \& {van den Heuvel}}{2006}]{2006csxs.book..623T}
{Tauris} T.~M.,  {van den Heuvel} E.~P.~J.,  2006, in {Lewin} W. H.~G.,  {van der Klis} M.,  eds, , Vol.~39, Compact stellar X-ray sources.
Cambridge University Press, pp 623--665, \mn@doi{10.48550/arXiv.astro-ph/0303456}

\bibitem[\protect\citeauthoryear{{Thomas}, {Gudennavar}  \& {Bubbly}}{{Thomas} et~al.}{2023}]{2023MNRAS.525.2355T}
{Thomas} N.~T.,  {Gudennavar} S.~B.,   {Bubbly} S.~G.,  2023, \mn@doi [mnras] {10.1093/mnras/stad2379}, \href {https://ui.adsabs.harvard.edu/abs/2023MNRAS.525.2355T} {525, 2355}

\bibitem[\protect\citeauthoryear{Titarchuk, Lapidus  \& Muslimov}{Titarchuk et~al.}{1998}]{Titarchuk_1998}
Titarchuk L.,  Lapidus I.,   Muslimov A.,  1998, \mn@doi [The Astrophysical Journal] {10.1086/305642}, 499, 315

\bibitem[\protect\citeauthoryear{{Titarchuk}, {Bradshaw}, {Geldzahler}  \& {Fomalont}}{{Titarchuk} et~al.}{2001}]{2001ApJ...555L..45T}
{Titarchuk} L.~G.,  {Bradshaw} C.~F.,  {Geldzahler} B.~J.,   {Fomalont} E.~B.,  2001, \mn@doi [apjl] {10.1086/323160}, \href {https://ui.adsabs.harvard.edu/abs/2001ApJ...555L..45T} {555, L45}

\bibitem[\protect\citeauthoryear{{Ursini, F.} et~al.,}{{Ursini, F.} et~al.}{2023}]{Ursini2023}
{Ursini, F.} et~al., 2023, \mn@doi [A&A] {10.1051/0004-6361/202346541}, 676, A20

\bibitem[\protect\citeauthoryear{{Ursini} et~al.,}{{Ursini} et~al.}{2023}]{2023A&A...676A..20U}
{Ursini} F.,  et~al., 2023, \mn@doi [aap] {10.1051/0004-6361/202346541}, \href {https://ui.adsabs.harvard.edu/abs/2023A&A...676A..20U} {676, A20}

\bibitem[\protect\citeauthoryear{{Ursini} et~al.,}{{Ursini} et~al.}{2024}]{2024Galax..12...43U}
{Ursini} F.,  et~al., 2024, \mn@doi [Galaxies] {10.3390/galaxies12040043}, \href {https://ui.adsabs.harvard.edu/abs/2024Galax..12...43U} {12, 43}

\bibitem[\protect\citeauthoryear{{Weisskopf} et~al.,}{{Weisskopf} et~al.}{2022}]{2022JATIS...8b6002W}
{Weisskopf} M.~C.,  et~al., 2022, \mn@doi [Journal of Astronomical Telescopes, Instruments, and Systems] {10.1117/1.JATIS.8.2.026002}, \href {https://ui.adsabs.harvard.edu/abs/2022JATIS...8b6002W} {8, 026002}

\bibitem[\protect\citeauthoryear{{White}, {Stella}  \& {Parmar}}{{White} et~al.}{1988}]{1988ApJ...324..363W}
{White} N.~E.,  {Stella} L.,   {Parmar} A.~N.,  1988, \mn@doi [apj] {10.1086/165901}, \href {https://ui.adsabs.harvard.edu/abs/1988ApJ...324..363W} {324, 363}

\bibitem[\protect\citeauthoryear{Wijnands \& van~der Klis}{Wijnands \& van~der Klis}{1999}]{Wijnands_1999}
Wijnands R.,  van~der Klis M.,  1999, \mn@doi [The Astrophysical Journal] {10.1086/307698}, 522, 965

\bibitem[\protect\citeauthoryear{{Zdziarski}, {Johnson}  \& {Magdziarz}}{{Zdziarski} et~al.}{1996}]{1996MNRAS.283..193Z}
{Zdziarski} A.~A.,  {Johnson} W.~N.,   {Magdziarz} P.,  1996, \mn@doi [mnras] {10.1093/mnras/283.1.193}, \href {https://ui.adsabs.harvard.edu/abs/1996MNRAS.283..193Z} {283, 193}

\bibitem[\protect\citeauthoryear{{van den Berg} \& {Homan}}{{van den Berg} \& {Homan}}{2017}]{2017ApJ...834...71V}
{van den Berg} M.,  {Homan} J.,  2017, \mn@doi [\apj] {10.3847/1538-4357/834/1/71}, \href {https://ui.adsabs.harvard.edu/abs/2017ApJ...834...71V} {834, 71}

\bibitem[\protect\citeauthoryear{van~der Klis}{van~der Klis}{1989}]{annurev.aa.27.090189.002505}
van~der Klis M.,  1989, \mn@doi [Annual Review of Astronomy and Astrophysics] {https://doi.org/10.1146/annurev.aa.27.090189.002505}, 27, 517

\bibitem[\protect\citeauthoryear{{van der Klis}}{{van der Klis}}{1995}]{1995xrbi.nasa..252V}
{van der Klis} M.,  1995, in {Lewin} W. H.~G.,  {van Paradijs} J.,   {van den Heuvel} E. P.~J.,  eds, X-ray Binaries. pp 252--307

\bibitem[\protect\citeauthoryear{{van der Klis}}{{van der Klis}}{2006}]{2006csxs.book...39V}
{van der Klis} M.,  2006, in {Lewin} W. H.~G.,  {van der Klis} M.,  eds, , Vol.~39, Compact stellar X-ray sources.
Cambridge University Press, pp 39--112

\makeatother
\end{thebibliography}

\bsp	
\label{lastpage}
\end{document}